\documentclass[3p]{elsarticle}

\newcommand{\ket}[1]{| #1 \rangle}
\newcommand{\bra}[1]{\langle #1 |}
\newcommand{\Rho}{\mathrm{P}}
\newcommand{\tr}[1]{\mathrm{tr}{\left\{#1\right\}}}
\newcommand{\Exp}[1]{\mathrm{e}^{\mbox{\footnotesize$#1$}}}
\newcommand{\I}{\mathrm{i}}

\makeatletter
\renewcommand{\ps@pprintTitle}{%
\renewcommand{\@oddhead}{\@empty}
\renewcommand{\@oddfoot}{\small%
\hfill Posted on the arXiv on 24 July 2009}%
\renewcommand{\@evenhead}{\@oddhead}
\renewcommand{\@evenfoot}{\@oddfoot}
}

\newcommand{\ps@RunTitle}{%
\renewcommand{\@oddfoot}{\@empty}
\renewcommand{\@evenfoot}{\@empty}
\renewcommand{\@oddhead}{%
{\underline{\makebox[\textwidth][s]%
{\footnotesize Teo,~Zhu,~Englert:~Qubit-pair~tomography\hfil\thepage}}}}
\renewcommand{\@evenhead}{\@oddhead}
}
\makeatother

\begin{document}

\title{%
Product measurements and fully symmetric measurements\newline 
in qubit-pair tomography: A numerical study\tnoteref{KW}}

\tnotetext[KW]{We dedicate this work to the memory of Krzysztof
  W\'odkiewicz, a friend and colleague who left us much too early.}

\author[cqt,phy,ngs]{\textsc{Teo} Yong Siah}

\author[cqt,ngs]{\textsc{Zhu} Huangjun}

\author[cqt,phy]{Berthold-Georg \textsc{Englert}}

\address[cqt]{Centre for Quantum Technologies, %
National University of Singapore, Singapore 117543, Singapore}

\address[phy]{Department of Physics, National University of Singapore, 
Singapore 117542, Singapore}

\address[ngs]{NUS Graduate School for Integrative Sciences and Engineering,
Singapore 117597, Singapore}

\begin{abstract}
State tomography on qubit pairs is routinely carried out by measuring the two
qubits separately, while one expects a higher efficiency from tomography with
highly symmetric joint measurements of both qubits.
Our numerical study of simulated experiments does not support such
expectations. 
\end{abstract}


\maketitle

\section{Introduction}\label{sec:intro}
Quantum state tomography makes efficient use of the resources if it employs a
probability operator measurement (POM) with the minimal number of outcomes ---
${4=2^2}$ for single-qubit tomography, ${16=4^2}$ for qubit-pair tomography,
$d^2$ for tomography in a $d$-dimensional Hilbert space.
Ideally, one would like to acquire tomographic data with the aid of a
symmetric informationally complete (SIC) measurement 
\cite{Zauner:99,Renes+3:04} 
because it probes the state space in an unbiased way and promises minimal
statistical errors on average when many, uniformly distributed, quantum states
are measured~\cite{Rehacek+2:04,Appleby+2:07}.

The SIC tetrahedron measurement \cite{Rehacek+2:04,Englert+3:05} has been
implemented for polarization qubits of single photons
\cite{Ling+3:06a,Ling+3:06b}, and the product POM that is composed of the
joint measurements of two single-qubit SIC~POMs has been used successfully for
two-qubit state tomography~\cite{Ling+2:08}.  
This product POM is, however, not a SIC~POM for the qubit pair, which invites
the question whether the realization of the two-qubit SIC~POM, with its
challenge of implementing projections to entangled states, is worth the
effort. 
Our answer is: Hardly. 

The evidence that justifies this answer is numerical.
We generate simulated tomographic data, both for the product POM and the
SIC~POM, for two-qubit states that are chosen either specifically or randomly.
Then we estimate the state from the simulated data and compare the
reconstructed state with the pre-chosen true state.
The figure of merit is the trace-class distance between the two states.

The paper is organized as follows.
We introduce notation and terminology in Sec.~\ref{sec:schemes} where we
briefly review schemes for state tomography.
Then, in Sec.~\ref{sec:gpsic}, we remark on group covariant SIC~POMs and
specify the product POM and SIC~POM that we use for the
numerical study. 
Its ingredients --- state estimation and the generation of random samples of
states --- are described in Secs.~\ref{sec:recon} and \ref{sec:random},
respectively. 
The results of the simulated experiments are presented in Sec.~\ref{sec:bell}
for tomography of maximally entangled states, and in Sec.~\ref{sec:average}
for averages over random samples of states.
We close with offering our conclusion in Sec.~\ref{sec:sum}.

\section{Measurement schemes for state tomography}
\label{sec:schemes} 
A POM\footnote{POM, with its emphasis on ``probability'' and ``measurement'' is
  quantum physics jargon. The corresponding popular mathematical term POVM
  (positive-operator valued measure) makes reference to measure theory.}  
is a decomposition of the identity into $J$ probability operators $\Pi_j$,
which we call \emph{outcomes},
\begin{equation}
  \label{eq:POMdef}
  \sum_{j=0}^{J-1}{\Pi_j}=1\qquad\mbox{with}\quad \Pi_j>0\,.
\end{equation}
The probabilities
\begin{equation}
  \label{eq:probs}
  p_j=\tr{\rho\Pi_j}\qquad\mbox{with}\quad\sum_{j=0}^{J-1}p_j=1
\end{equation}
permit a reconstruction of the statistical operator $\rho$, the \emph{state}
of the physical system, if the POM is informationally complete (IC).
In a $d$-dimensional Hilbert space, an IC~POM must have at least $d^2$
outcomes, and a minimal IC~POM has exactly ${J=d^2}$ outcomes and no redundant
ones, and then the state $\rho$ is uniquely expressed in terms of the
probabilities $p_j$ and the \emph{reconstruction operators} $\Rho_j$ 
that are dual to the outcomes $\Pi_j$,
\begin{equation}
  \label{eq:reconstruct}
  \rho=\sum_{j=0}^{d^2-1}p_j\Rho_j\,,\qquad\mbox{where}\quad
  \tr{\Rho_j\Pi_k}=\delta_{jk}\,.
\end{equation}

A SIC~POM is a minimal IC~POM with the symmetry property that all outcomes
are unitarily equivalent with ``equal angles'' between each pair of
outcomes, 
\begin{eqnarray}
  \label{eq:gensic}
  \tr{\Pi_j\Pi_k}=a\delta_{jk}+b\qquad&\mbox{with}&\quad
  \tr{\Pi_j}=a+bd^2=\frac{1}{d}\nonumber\\ 
 &\mbox{and}&\quad \frac{1}{d^3}<\tr{\Pi_j^2}=a+b\leq\frac{1}{d^2}\,.
\end{eqnarray}
The reconstruction operators for such a SIC~POM are
\begin{equation}
  \label{eq:reconop}
  \Rho_j=\frac{1}{a}\bigl(\Pi_j-bd\bigr)\,,
\end{equation}
so that
\begin{equation}
  \label{eq:reconop1}
  \tr{\Rho_j}=1\,,\quad\tr{\Rho_j\Rho_k}=\frac{1}{a}(\delta_{jk}-bd)\,,\quad
\sum_{j=0}^{d^2-1}P_j=d\,.
\end{equation}
These translate $d^{-1}\leq\tr{\rho^2}\leq1$ into
\begin{equation}
  \label{eq:purity}
  \sum_{j=0}^{d^2-1}\Bigl(p_j-\frac{1}{d^2}\Bigr)^2\leq\frac{d-1}{d}a\,,
\end{equation}
a second constraint obeyed by the probabilities $p_j$ in addition to the unit
sum in (\ref{eq:probs}).
The equal sign in (\ref{eq:purity}) applies when $\rho$ is a pure state,
${\rho^2=\rho}$, and only then.

When the $d^2$ outcomes $\Pi_j$ of a SIC~POM are rank-$1$ operators, 
that is: when they are subnormalized projectors to pure states, 
then ${a+b}$ assumes the maximal value permitted by (\ref{eq:gensic}),
\begin{eqnarray}
  \label{eq:puresic}
  &\textrm{rank-$1$ SIC~POM:}\quad& 
a+b=\frac{1}{d^2}\,,\quad a=bd=\frac{1}{d+d^2}\,,\nonumber\\
&& \Pi_j^2=\frac{1}{d}\Pi_j^{\ }\,,\quad \Rho_j^{\ }=(d+d^2)\Pi_j^{\ }-1\,.
\end{eqnarray}
Almost all SIC~POMs studied in the published literature are of
this rank-$1$ kind. 
One exception is the two-qubit SIC~POM of Ref.~\cite{Zhu+2:09} whose rank-$3$
outcomes are optimal entanglement witnesses. 
In the present context, we will only deal with standard rank-$1$ SIC~POMs.

\section{Group covariant SIC~POMs and their fiducial states} 
\label{sec:gpsic}
As conjectured in \cite{Zauner:99,Renes+3:04}, for every Hilbert space
dimension $d$, there exist rank-$1$ SIC~POMs which are covariant with respect
to the Heisenberg--Weyl group, also called the generalized Pauli group
because it is the $d$-dimensional analog of the ${d=2}$ group of unitary
transformations effected by the Pauli operators $\sigma_x$, $\sigma_z$ and
their products, including $\sigma_x\sigma_z=-\I\sigma_y$ as well as the
identity ${\sigma_x^2=1}$. 
Indeed, SIC~POMs of this group covariant kind are typical.

The Heisenberg--Weyl group refers to a pre-chosen basis of kets $\ket{0}$,
$\ket{1}$, \dots, $\ket{d-1}$.
They are eigenkets of the $d$-periodic unitary operator $Z$, the analog of
$\sigma_z$,
\begin{equation}
  \label{eq:HW1}
  Z\ket{k}=\ket{k}\Exp{\I2\pi k/d}\,,\qquad Z^d=1\,,
\end{equation}
and are cyclically permuted by the $d$-periodic unitary
operator $X$, the analog of $\sigma_x$,
\begin{equation}
  \label{eq:HW2}
  X\ket{k}=\ket{k+1}\quad\mbox{for $k=0,1,\dots,d-2$}\,,\quad
  X\ket{d-1}=\ket{0}\,,\qquad X^d=1\,.
\end{equation}
The $(m,n)$th element in the abelian Heisenberg--Weyl group of unitary
transformations is the mapping
\begin{eqnarray}
  \label{eq:HW3}
 & F \to X^mZ^nFZ^{-n}X^{-m}=Z^nX^mFX^{-m}Z^{-n}&\nonumber\\
 &\mbox{for $m,n=0,1,\dots,d-1$}\,,&
\end{eqnarray}
where $F$ is any linear operator on the $d$-dimensional Hilbert space of kets.
Although $X$ and $Z$ do not commute, $ZX=\Exp{\I2\pi/d}XZ$, the order of
factors does not matter in the transformation (\ref{eq:HW3}), and so we have
$d^2$ group elements.

A group covariant SIC~POM is composed of the outcomes
\begin{equation}
  \label{eq:HW4}
  \Pi_{mn}= X^mZ^n \Pi_{00}Z^{-n}X^{-m}\,,
\end{equation}
where the $d^2$ pairs $m,n$ take over the role of label $j$ in
(\ref{eq:gensic}) and $\Pi_{00}$ is the \emph{fiducial outcome}.
The ergodic property of the Heisenberg--Weyl group,
\begin{equation}
  \label{eq:HW5}
  \sum_{m,n=0}^{d-1} X^mZ^nFZ^{-n}X^{-m}=d\,\tr{F}\,,
\end{equation}
ensures that the $\Pi_{mn}$s are proper outcomes of a POM if $\Pi_{00}$ is a
probability operator with ${\tr{\Pi_{00}}=1/d}$.
But the rank-$1$ SIC~POM condition of (\ref{eq:gensic}) with $a$ and $b$ from
(\ref{eq:puresic}) is obeyed only if 
$\Pi_{00}=\ket{\mathrm{f}}d^{-1}\bra{\mathrm{f}}$ is such that
\begin{equation}
  \label{eq:HW6}
  \bigl|\bra{\mathrm{f}} X^mZ^n \ket{\mathrm{f}}\bigr|^2
=\frac{1+d\,\delta_{m0}\,\delta_{n0}}{1+d}\quad
\mbox{for $m,n=0,1,\dots,d-1$}\,,
\end{equation}
where the normalized $\ket{\mathrm{f}}$ is the fiducial ket or
\emph{seed}~\cite{Renes+3:04}. 
The set of equations (\ref{eq:HW6}) can serve as the basis for a numerical
search, successfully completed by Renes \textit{et al.} up to
${d=45}$~\cite{Renes+3:04}, but there are also analytically known fiducial
kets and their rank-$1$ POMs.

For the qubit case ${d=2}$, we have the familiar 
\emph{tetrahedron measurement} with the outcomes
\begin{eqnarray}
  \label{eq:tetra1}
  T_0&=&\frac{1}{4\sqrt{3}}\bigl(\sqrt{3}+\sigma_x+\sigma_y+\sigma_z\bigr)\,,
\nonumber\\
  T_1&=&\frac{1}{4\sqrt{3}}\bigl(\sqrt{3}+\sigma_x-\sigma_y-\sigma_z\bigr)
        =\sigma_xT_0\sigma_x\,,
\nonumber\\
  T_2&=&\frac{1}{4\sqrt{3}}\bigl(\sqrt{3}-\sigma_x-\sigma_y+\sigma_z\bigr)
        =\sigma_zT_0\sigma_z\,,
\nonumber\\
  T_3&=&\frac{1}{4\sqrt{3}}\bigl(\sqrt{3}-\sigma_x+\sigma_y-\sigma_z\bigr)
        =\sigma_x\sigma_zT_0\sigma_z\sigma_x\,,
\end{eqnarray}
for which
\begin{equation}
  \label{eq:tetra2}
  \sum_{j=0}^3T_j=1\,,\quad\tr{T_j}=\frac{1}{2}\,,\quad
  \tr{T_jT_k}=\frac{1}{6}\delta_{jk}+\frac{1}{12}\,.
\end{equation}
Its name derives from the geometry of the four Pauli vectors $\vec{t}_j$,
\begin{equation}
  \label{eq:tetra3}
  T_j=\frac{1}{4}\bigl(1+\vec{t}_j\cdot\vec{\sigma}\bigr)\,,\qquad
  \vec{t}_j\cdot\vec{t}_k=\frac{4}{3}\delta_{jk}-\frac{1}{3}\,, 
\end{equation}
whose tips are the corners of a tetrahedron inside the Bloch sphere.
See Ref.~\cite{Rehacek+2:04} for the properties of the tetrahedron
measurement, and Ref.~\cite{Ling+3:06a} for its implementation for
experimental tomography of photon polarization qubits.

The tensor product of two tetrahedron SIC~POMs,
\begin{equation}
  \label{eq:tetra4}
  T_{mn}=T_m\otimes T_n\,,\qquad \sum_{m,n=0}^3T_{mn}=1\,,
\end{equation}
is a perfectly suitable POM for two-qubit tomography.
It is central to the Singapore protocol for quantum key distribution
\cite{Englert+5:04}, and has been used in the experiments by Ling
\textit{et al.}~\cite{Ling+2:08}. 
This product POM is, however, not a SIC~POM for the qubit pair.

For two-qubit SIC tomography, we can use any one of the known
SIC~POMs for ${d=4}$, but they are not equivalent in their two-qubit
properties, such as the concurrence of the outcomes~\cite{Zhu+2:ip}.
We prefer a particular SIC~POM, generated from Appelby's fiducial
state~\cite{Appleby:05},
\begin{equation}
  \label{eq:appleseed}
  \ket{\mathrm{f}_{\mathrm{A}}}=\bigl(\ket{00},\ket{01},\ket{10},\ket{11}\bigr)
\frac{1}{2\sqrt{3+G}}\left(\begin{array}{c}
1+\Exp{-\I\pi/4} \\[1ex] \Exp{\I\pi/4}+\I G^{-3/2}\\[1ex]    
1-\Exp{-\I\pi/4} \\[1ex] \Exp{\I\pi/4}-\I G^{-3/2}
\end{array}\right),
\end{equation}
where $G=(\sqrt{5}-1)/2$ is the golden ratio.
The transformations of the Heisenberg--Weyl group refer to the two-qubit kets
${\ket{j_1j_2}=\ket{j_1}\otimes\ket{j_2}}$ in the order stated, so that we
have
\begin{equation}
  \label{eq:2qubitHW}
  Z=\frac{1+\I}{2}\sigma_z\otimes(1-\I\sigma_z)\,,\quad
  X=\frac{1}{2}(1+\sigma_x)\otimes\sigma_x
    -\frac{\I}{2}(1-\sigma_x)\otimes\sigma_y
\end{equation}
for the two-qubit versions of the basic unitary operators 
in (\ref{eq:HW1}) and (\ref{eq:HW2}).

The outcomes $\Pi_{mn}$ of the particular SIC~POM that we get from the
seed (\ref{eq:appleseed}),
\begin{equation}
  \label{eq:applePOM}
  \Pi_{mn}=X^mZ^n\ket{\mathrm{f}_{\mathrm{A}}}\frac{1}{d}
          \bra{\mathrm{f}_{\mathrm{A}}}Z^{-n}X^{-m}
  \qquad\mbox{for $m,n=0,1,2,3$}\,,
\end{equation}
possess an additional symmetry: all the $16$ two-qubit states $d\,\Pi_{mn}$ have
the same concurrence of $\sqrt{2/5}$.
These two-qubit states are typical in the sense that their squared
concurrence equals the average squared concurrence of all pure two-qubit
states~\cite{Zyczkowski+1:01}.
As a consequence of the common concurrence, 
one can turn the $\Pi_{mn}$s into each other with the aid of
\emph{local} unitary transformations that act on one of the qubits only.
This is a rather peculiar feature, not shared by other SIC~POMs for qubit
pairs~\cite{Zhu+2:ip}.

\section{Techniques for state estimation}
\label{sec:recon}
The data acquired in the simulated experiments are simply the counts of clicks
of each of the $16$ detectors, one for each outcome of the two-qubit POMs ---
the product POM of (\ref{eq:tetra4}) and the SIC~POM of (\ref{eq:appleseed})
and (\ref{eq:applePOM}). 
Owing to the statistical fluctuations that originate in the quantum
indeterminism, the relative frequencies of detector counts are not equal to
the probabilities $p_{mn}^{\mathrm{(prod)}}=\tr{\rho T_{mn}}$ and
$p_{mn}^{\mathrm{(SIC)}}=\tr{\rho\Pi_{mn}}$, at best the respective relative
frequencies $\tilde{p}_{mn}^{\mathrm{(prod)}}$ and $\tilde{p}_{mn}^{\mathrm{(SIC)}}$
approximate the intrinsic probabilities.

Therefore, there is danger in relying on (\ref{eq:reconstruct}).
The reconstructed states
\begin{eqnarray}  \label{eq:rho-estim}
\rho_{\mathrm{est}}^{\mathrm{(prod)}}
&=&\sum_{m,n=0}^{3}\tilde{p}_{mn}^{\mathrm{(prod)}}(6T_m-1)\otimes(6T_n-1)\,,
\nonumber\\
\rho_{\mathrm{est}}^{\mathrm{(SIC)}}
&=&\sum_{m,n=0}^{3}\tilde{p}_{mn}^{\mathrm{(SIC)}}(20\,\Pi_{mn}-1)
\end{eqnarray}
are not guaranteed to be positive operators.
It is possible, and indeed happens regularly when the true state has high
purity, that these \emph{raw data} (RD) estimates have negative eigenvalues 
because the relative frequencies do not obey restrictions such as the
appropriate version of (\ref{eq:purity}).
In the context of single-qubit tomography with the tetrahedron measurement,
this matter is discussed in section IV~C of Ref.~\cite{Rehacek+2:04}.
A thorough coverage of many aspects of quantum state estimation is given
in the 2004 monograph of Paris and \v{R}eh\'a\v{c}ek~\cite{LNP649}.  

The probability that a total of $N$ detector clicks is recorded with the
actual relative frequencies $\tilde{p}_{mn}$ is the
\emph{likelihood} $\mathcal{L}(\rho)$. 
Its logarithm is
\begin{equation}
  \label{eq:L}
  \log\bigl(\mathcal{L}(\rho)\bigr)=N\sum_{m,n=0}^3\tilde{p}_{mn}\log p_{mn}\,,
\end{equation}
where the unknown intrinsic probabilities $p_{mn}$ derive from the true
statistical operator $\rho$ that we wish to estimate.
In the \emph{maximum likelihood} (ML) strategy we take the $\rho$ for which
$\mathcal{L}(\rho)$ is largest as our best guess for the true $\rho$,
\begin{equation}
  \label{eq:ML}
  \mathcal{L}(\rho_{\mathrm{est}})=\max\limits_{\rho}\mathcal{L}(\rho)\,.
\end{equation}
The ML estimate is identical with the respective RD estimate of
(\ref{eq:rho-estim}) if the RD estimate is permissible, that is: if there is a
physical $\rho$ whose probabilities are equal to the recorded relative
frequencies. 
Accordingly, we need to calculate the ML estimate only if the RD
reconstruction (\ref{eq:rho-estim}) fails.

The ML estimate is found by a numerical iteration that begins with an
arbitrary guess $\rho_{\mathrm{est}}^{(0)}$, for which the completely mixed
state $\rho_{\mathrm{est}}^{(0)}=1/4$ is an option, and then determines
improved estimates successively in accordance with the update rule
\begin{equation}
  \label{eq:MLupdate}
  \rho^{(k+1)}_{\mathrm{est}}
=\frac{\bigl[1+\epsilon_kR_k\bigr]\rho_{\mathrm{est}}^{(k)}
       \bigl[1+\epsilon_kR_k\bigr]}
{\tr{\bigl[1+\epsilon_kR_k\bigr]\rho_{\mathrm{est}}^{(k)}
       \bigl[1+\epsilon_kR_k\bigr]}}\,,
\end{equation}
where (replace $\Pi_{mn}$ by $T_{mn}$ for the product POM)
\begin{equation}
  \label{eq:MLR}
  R_k=\sum_{m,n=0}^3\frac{\tilde{p}_{mn}\Pi_{mn}}
                         {\tr{\rho_{\mathrm{est}}^{(k)}\Pi_{mn}}}-1
\end{equation}
and ${\epsilon_k>0}$ is a suitably chosen increment (see below).
The iteration stops when the trace class distance 
$\tr{\bigl|R_k\rho_{\mathrm{est}}^{(k)}\bigr|}$ is
below the pre-set accuracy threshold.
The rule (\ref{eq:MLupdate}), which is a variant of the recipe given in
chapter~3 of \cite{LNP649}, is such that all $\rho^{(k)}_{\mathrm{est}}$s are
assuredly positive and normalized to unit trace.

In a bit of detail, the important steps of the iteration are as follows.
\begin{quotation}\noindent%
Given $\rho_{\mathrm{est}}^{(k)}$,
\begin{enumerate}\renewcommand{\theenumi}{\textbf{\roman{enumi}}}
 \item Compute $R_k$; terminate the iteration if
       $\tr{\bigl|R_k\rho_{\mathrm{est}}^{(k)}\bigr|}$
               is below the accuracy threshold, 
               otherwise proceed.
 \item Use two trial values for $\epsilon_k$ to compute two
       $\rho_{\mathrm{est}}^{(k+1)}$s and determine the value of the likelihood 
       $\mathcal{L}\bigl(\rho_{\mathrm{est}}^{(k+1)}\bigr)$
       for both.  
 \item Combine these two with the $\epsilon_k=0$ value  
       $\mathcal{L}\bigl(\rho_{\mathrm{est}}^{(k)}\bigr)$
       and compute a quadratic function of $\epsilon_k$ that interpolates
       between the three support values.
 \item Find the $\epsilon_k$ value for which the quadratic function assumes
       its maximum.
 \item Use this maximizing $\epsilon_k$ in (\ref{eq:MLupdate}) for the update 
       $\rho_{\mathrm{est}}^{(k)}\to\rho_{\mathrm{est}}^{(k+1)}$.
 \item Repeat.
\end{enumerate}
\end{quotation}
With the optimization of $\epsilon_k$ in steps \textbf{ii}--\textbf{iv}, 
convergence is quite fast in practice.

\section{Choosing states at random}
\label{sec:random}
For the simulated state tomography, we choose the true states either
specifically or at random. 
We generate the random states by a procedure described in
Refs.~\cite{Zyczkowski+1:01,BengtssonZyczkowski}.  
A random rank-$r$ state $\rho$ is selected by first populating the entries of
a $4\times r$ matrix $Y$ such that the real and imaginary parts follow 
noncentral Gaussian distributions offset by mean values given by the matrix
$M=\mathrm{E}(Y)$. 
Then 
\begin{equation}
  \label{eq:randrho}
  \rho=\bigl(\ket{00},\ket{01},\ket{10},\ket{11}\bigr)
       \frac{YY^\dagger}{\tr{YY^\dagger}}
       \left(\begin{array}{c}
           \bra{00}\\ \bra{01}\\ \bra{10}\\ \bra{11}
         \end{array}\right)
\end{equation}
is a rank-$r$ state in the random sample that is characterized by matrix $M$.
The choice $M=0$ for $r=4$ gives a uniform sample with respect to the
Hilbert--Schmidt distance, by other choices we can regulate the average purity
of the states in the random sample. 

Random samples of pure states (${r=1}$) are unitarily invariant and thus
uniform. 
So are random samples of unbiased mixed states (${M=0}$),
but if we choose ${M\neq0}$ for a mixed states sample, there will be a bias
because we lose the unitary equivalence. 
The average performance of the SIC~POM will depend on the sample or,
equivalently, which one of the unitarily equivalent SIC~POMs we use for the
same biased sample.

Once the true state $\rho$ is chosen for the simulation, we use the resulting
quantum probabilities in conjunction with a standard (pseudo-)random number
generator to get a total of $N$ simulated detector clicks.
The relative frequencies $\tilde{p}_{mn}$ of detector clicks then determine
the RD estimate of $\rho$ and, if need be, also the ML estimate.
The quality of the estimate is judged by the trace-class distance,
\begin{equation}
  \label{eq:distance}
  D=\frac{1}{2}\tr{\bigl|\rho-\rho_{\mathrm{est}}\bigr|}\,,
\end{equation}
with ${0\leq D\leq1}$. 
We have $D=0$ only if the two states are equal, ${\rho_{\mathrm{est}}=\rho}$, 
and the maximal distance ${D=1}$ when they are orthogonal 
${\rho_{\mathrm{est}}\rho=0}$.
Typically, the distance decreases ${D\propto N^{-1/2}}$ as the sample size $N$
increases.

\section{Tomography of maximally entangled pure states}
\label{sec:bell}

\begin{figure}[tp]
\centerline{\includegraphics[bb=92 231 449 747,clip=]{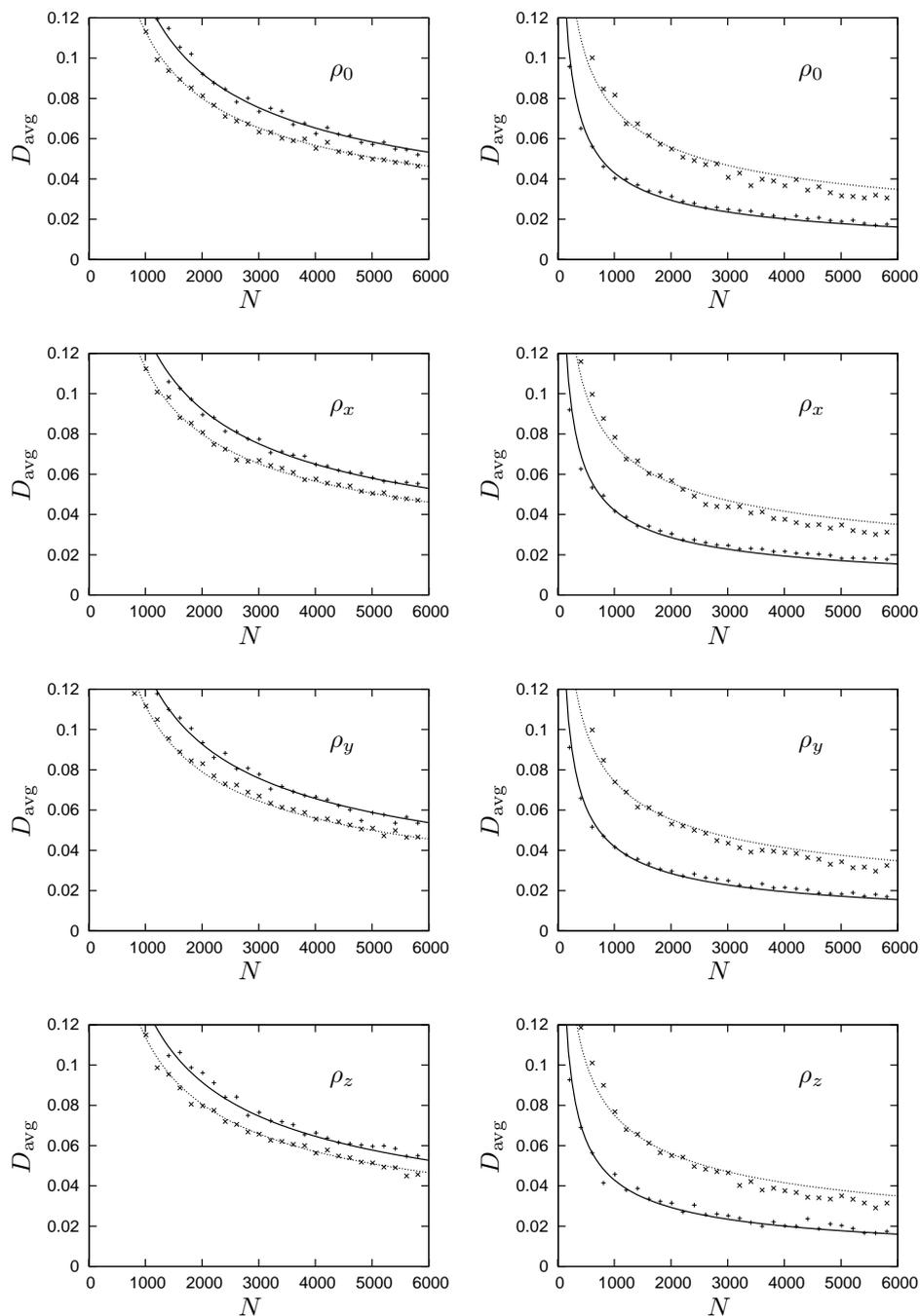}}
\caption{\label{fig:bellstatetomo}%
  Simulated tomography of two-qubit Bell states
  with the product POM (crosses $+$ and solid
  line) or the  SIC~POM (crosses $\times$ and dotted line) with raw-data 
  state estimation (left-side plots) and maximum-likelihood state estimation 
  (right-side plots).
  The trace-class distance $D_{\mathrm{avg}}$ is averaged over a hundred
  simulation runs with up to $6000$ measured qubit pairs, all prepared in the
  respective Bell state.
  The solid and dotted lines are least-square 
  fits to a power law ${D_{\mathrm{avg}}=a/N^c}$.}
\end{figure}

We now compare the tomographic performances of RD and ML reconstruction for
pure states for which the failure of RD estimation is typical rather than
exceptional.
In particular, we choose as the true states the four standard 
Bell states that are given by 
\begin{eqnarray}\label{eq:rhoBell}
\rho_0&=&\frac{1}{2}\bigl(1-\sigma_x\otimes\sigma_x
                           -\sigma_y\otimes\sigma_y
                           -\sigma_z\otimes\sigma_z\bigr)
       =\ket{\Psi^-}\bra{\Psi^-}\,,\nonumber\\
\rho_x&=&\frac{1}{2}\bigl(1-\sigma_x\otimes\sigma_x
                           +\sigma_y\otimes\sigma_y
                           +\sigma_z\otimes\sigma_z\bigr)
       =\ket{\Phi^-}\bra{\Phi^-}\,,\nonumber\\
\rho_y&=&\frac{1}{2}\bigl(1+\sigma_x\otimes\sigma_x
                           -\sigma_y\otimes\sigma_y
                           +\sigma_z\otimes\sigma_z\bigr)
       =\ket{\Phi^+}\bra{\Phi^+}\,,\nonumber\\
\rho_z&=&\frac{1}{2}\bigl(1+\sigma_x\otimes\sigma_x
                           +\sigma_y\otimes\sigma_y
                           -\sigma_z\otimes\sigma_z\bigr)
       =\ket{\Psi^+}\bra{\Psi^+}
\end{eqnarray}
with
\begin{equation}\label{eq:ketBell}
\ket{\Psi^\pm}=\frac{1}{\sqrt{2}}\bigl(\ket{01}\pm\ket{10}\bigr)
\quad\mbox{and}\quad
\ket{\Phi^\pm}=\frac{1}{\sqrt{2}}\bigl(\ket{00}\pm\ket{11}\bigr)\,.  
\end{equation}
The plots in the left column of Fig.~\ref{fig:bellstatetomo} show the $N$
dependence of the distance between the RD estimate and the respective Bell
state.
Not unexpectedly, the SIC~POM outperforms the product POM.
But this observation is, in fact, misleading as almost all RD estimates are
unphysical.
The comparison with the ML estimates in the right column is more to the point,
and there the picture is reversed: rather unexpectedly, the product~POM gives
better estimates than the SIC~POM. 

The simulated product-POM data compare well with the experimental data of
Ref.~\cite{Ling+2:08}, where a RD estimate was used.
The experimental values for the average trace-class distance $D_{\mathrm{avg}}$
showed a dependence $\propto N^{-1/2}$ on the total number of detector clicks,
and our simulated data follow the same reciprocal square-root law.

The striking similarity among the four plots of each column in
Fig.~\ref{fig:bellstatetomo} is not accidental but could in fact have been
anticipated. 
The transformations effected by the squares of the Heisenberg--Weyl
operators $X$ and $Z$ of
(\ref{eq:2qubitHW}) interchange the Bell states,
\begin{equation}
  \label{eq:transBell}
  X^2\rho_0X^2=\rho_x\,,\quad Z^2\rho_xZ^2=\rho_y\,,\quad
  X^2\rho_yX^2=\rho_z\,,\quad Z^2\rho_zZ^2=\rho_0\,,
\end{equation}
where ${X^2=\sigma_x\otimes1}$ and ${Z^2=1\otimes\sigma_z}$ act on one of the
qubits only.
It follows that the respective sets of probabilities $p_{mn}$ of the Bell
states are permutations of each other.

For a more systematic quantitative comparison between the product POM and the
SIC~POM, we introduce a \emph{performance factor} $\eta$, guided by the 
usual understanding of optimal tomography with SIC~POMs. 
Suppose the power laws have the form $D=a/N^c$, and 
$N^{\mathrm{(prod)}}$ and $N^{\mathrm{(SIC)}}$ are the respective number of
qubit pairs that need to be measured to reach the desired threshold value 
$D_{\mathrm{thr}}$. 
Then
\begin{equation}
  \label{eq:def-eta}
  \eta=\frac{N^{\mathrm{(prod)}}}{N^{\mathrm{(SIC)}}}
=\biggl(\frac{a^{\mathrm{(prod)}}}{D_{\mathrm{thr}}}
\biggr)^{\mbox{\footnotesize$1/c^{\mathrm{(prod)}}$}}
\biggl(\frac{D_{\mathrm{thr}}}{a^{\mathrm{(SIC)}}}
\biggr)^{\mbox{\footnotesize$1/c^{\mathrm{(SIC)}}$}}
\end{equation}
defines $\eta$.
For $\eta>1$, the larger its value, the faster the convergence rate of the
SIC~POM power law compared to that of the product POM for a fixed benchmark
distance $D_{\mathrm{thr}}$. 
Similarly, $\eta<1$ indicates that the product POM is more efficient.

From top to bottom, the $\eta$ values for the data in
Fig.~\ref{fig:bellstatetomo} are $1.32$, $1.31$, $1.37$, and $1.29$ for the RD
estimates of the left column, and $0.42$, $0.43$, $0.41$, and $0.42$ for the
ML estimates of the right column, all referring to the benchmark value
$D_{\mathrm{thr}}=0.1$. 
It is staggering how much better the product POM performs as soon as proper ML
estimation of the state is incorporated.
This observation is not limited to the maximally entangled Bell states.
We have seen it also for quite a few randomly chosen pure two-qubit states 
and for the three-qubit W state  
${\ket{\mathrm{W}}=\bigl(\ket{001}+\ket{010}+\ket{100}\bigr)/\sqrt{3}}$.
A more extensive study of three-quit tomography appears to be worthwhile.

That said, we must also note that the four Bell states of (\ref{eq:rhoBell})
are not typical among the maximally entangled states, owing to their particular
alignment with the outcomes of the two POMs under consideration.
When one averages over a large sample of maximally entangled states, see the
bottom right subtable in Table~\ref{tbl:averages}, the picture is different:
the average $\eta$ value is a bit larger than $1$.

\begin{table}[t]
\caption{Average values for three samples of $1000$ randomly chosen states each.
The top left table is for unbiased full-rank mixed states 
(${r=4}$ and ${M=0}$ in Sec.~\ref{sec:random});
the top right table is for biased full-rank mixed states 
(${r=4}$ and ${M\neq0}$)
whose purity $\tr{\rho^2}$ exceeds $0.8$ with high probability.
The bottom left table is for pure states (${r=1}$);
the bottom right table is for maximally entangled states.
In each table, the first and second rows report the average number of qubit
pairs that need to be measured before the benchmark value of
$D_{\mathrm{thr}}=0.1$ is reached, 
for raw-data state estimation (RD) and maximum-likelihood estimation (ML) as
well as for the product POM and for the SIC~POM.
The third rows show the performance factors $\eta$ of (\ref{eq:def-eta})
that compares the efficiency of the two POMs.   
For the more relevant ML estimation, there is no significant advantage of the
SIC~POM over the product POM, except perhaps for 
the sample of unbiased mixed states which is dominated by low-purity
states~\cite{BengtssonZyczkowski}.  
}
\begin{center}\vspace*{-\baselineskip}
\begin{tabular}{rl}
\begin{tabular}{ccc}
\multicolumn{3}{c}{\underline{\makebox[13em][c]{Unbiased mixed states}}%
\rule{0pt}{4ex}}
\\ \rule{0pt}{3ex}
     & RD & ML\\ \hline
prod & $1787\pm102$ & $1534\pm156$ \\
SIC  & $1349\pm67\ $  & $1165\pm101$ \\
$\eta$ & $1.33\pm0.09$ & $1.32\pm0.11$ \\
\hline\hline\end{tabular}&
\begin{tabular}{ccc}
\multicolumn{3}{c}{\underline{\makebox[13em][c]{Biased mixed states}}%
\rule{0pt}{4ex}}
\\ \rule{0pt}{3ex}
     & RD & ML\\ \hline
prod & $1739\pm114$ & $712\pm99$ \\
SIC  & $1293\pm67\ $  & $639\pm73$ \\
$\eta$ & $1.35\pm0.11$ & $1.12\pm0.16$ \\
\hline\hline\end{tabular}\\
\begin{tabular}{ccc}
\multicolumn{3}{c}{\underline{\makebox[13em][c]{Pure states}}%
\rule{0pt}{4ex}}
\\ \rule{0pt}{3ex}
     & RD & ML\\ \hline
prod & $1738\pm118$ & $472\pm73\ $ \\
SIC  & $1284\pm70\ $  & $455\pm108$ \\
$\eta$ & $1.35\pm0.16$ & $1.09\pm0.29$ \\
\hline\hline\end{tabular}&
\begin{tabular}{ccc}
\multicolumn{3}{c}{\underline{\makebox[13em][c]{Maximally entangled states}}%
\rule{0pt}{4ex}}
\\ \rule{0pt}{3ex}
     & RD & ML\\ \hline
prod & $1728\pm103$ & $552\pm89\ $ \\
SIC  & $1276\pm70\ $  & $442\pm72\ $ \\
$\eta$ & $1.36\pm0.11$ & $1.27\pm0.29$ \\
\hline\hline\end{tabular}
\end{tabular}\end{center}
\label{tbl:averages}
\end{table}

\section{Averaged performances}
\label{sec:average}
The data for average performances of the product POM and the SIC~POM for
randomly chosen states are collected in Table~\ref{tbl:averages}.
As explained in the table caption, we report data for four samples of random
states: an unbiased set of mixed states, a biased high-purity set of mixed
states, a set of pure states, and a sample of maximally entangled states.

The standard deviation for each quantity, 
as listed in Table~\ref{tbl:averages}, 
is a combination of two types of errors: one that comes
from performing a number of experiments on a fixed state and another incurred
as we sample the random states. 
To identify the two different errors, one would need to perform an extremely
large number of simulated experiments, in excess of $10,000$, on each state
and so reduce the former error to orders of magnitude smaller than the latter
error. 

One lesson of the numbers in Table~\ref{tbl:averages} is that the RD estimate
is not only sometimes unacceptable, but it also performs worse than the ML
estimate. 
Clearly, one should not rely at all on the RD estimate.

The second lesson is that there is no significant difference between the
efficiency of the SIC~POM and the product POM for states of high purity, and
there is not much of an advantage for the SIC~POM when the states have low
purity. 
The averages over samples of randomly chosen states do not confirm the
observation made for Bell states in Sec.~\ref{sec:bell}.

\section{Conclusion}
\label{sec:sum}
We have carried out a numerical study that compares the performance of the
product POM and the SIC~POM in qubit-pair tomography.
All evidence indicates that the product POM performs almost as well as the
SIC~POM; there is not much of an advantage in employing the SIC~POM.
Since it is much more challenging to realize the SIC~POM in a laboratory
experiment than the product POM, we conclude that the implementation of the
SIC~POM for tomographic purposes is hardly worth the trouble.

\section*{Acknowledgments}
Centre for Quantum Technologies is a Research Centre of Excellence funded by
Ministry of Education and National Research Foundation of Singapore.

\newcommand{\arXiv}[2][quant-ph]{\textrm{eprint\ arXiv:#1/#2}}
\newcommand{\arxiv}[2][quant-ph]{\textrm{eprint\ arXiv:#2 [#1]}}

\end{document}